\documentclass[a4paper,11pt]{article}
\usepackage{bm,mathrsfs,amsmath,amsthm,amssymb,amscd,longtable,array,graphicx,color}

\newcommand{\nc}{\newcommand}
\nc{\rnc}{\renewcommand}
\nc{\nn}{\nonumber}
\nc{\der}{{\partial}}
\rnc{\Im}{{\rm{Im}\,}}
\rnc{\Re}{{\rm{Re}\,}}
\nc{\db}{\displaybreak[0]\\}
\nc{\bra}{\langle}
\nc{\ket}{\rangle}
\nc{\bs}{\boldsymbol}

\newtheorem{theorem}{Theorem}[section]

\newtheorem{proposition}[theorem]{Proposition}

\theoremstyle{definition}

\newtheoremstyle{query}%
{}{}
{\color{red}}
{}
{\sffamily\bfseries}{:}{12pt}
{}
\theoremstyle{query}

\numberwithin{equation}{section}

\numberwithin{equation}{section}

\begin{document}%
%
\title{Elliptic free-fermion model with OS boundary \\
and elliptic Pfaffians}

\author{
Kohei Motegi\thanks{E-mail: kmoteg0@kaiyodai.ac.jp}
\\\\
{\it Faculty of Marine Technology, Tokyo University of Marine Science and Technology,}\\
 {\it Etchujima 2-1-6, Koto-Ku, Tokyo, 135-8533, Japan} \\
\\\\
\\
}

\date{\today}

\maketitle

\begin{abstract}
We introduce and study a class of partition functions of 
an elliptic free-fermionic face model.
We study the partition functions with a triangular boundary
using the off-diagonal $K$-matrix at the boundary
(OS boundary), which was introduced by
Kuperberg as a class of variants
of the domain wall boundary partition functions.
We find explicit forms 
of the partition functions with OS boundary
using elliptic Pfaffians.
We find two expressions based on two versions
of Korepin's method,
and we obtain an identity between two elliptic Pfaffians as a corollary.

\end{abstract}

\section{Introduction}
Special kinds of determinants and Pfaffians are not only interesting
on their own but also because of their appearances in
many fields of mathematics and mathematical physics.
In mathematical physics, they often appear as partition functions
of integrable lattice models
\cite{Bethe,FST,Baxter,KBI,JM,Re}.
Among the most notable examples are the works by Korepin and Izergin.
Korepin \cite{Ko} introduced the domain wall boundary partition functions (DWBPF)
of the $U_q(sl_2)$ six-vertex model,
and also introduced a technique which enables one to reduce the
problem of finding the explicit forms of the DWBPF to
finding polynomials which satisfy several properties which
uniquely define them. Later, Izergin \cite{Iz} found the explicit determinant form
which is now called the Izergin-Korepin determinant.
Several variants of the domain wall boundary partition functions
were introduced and studied, sometimes with applications to the enumeration
of the alternating sign matrices and connections with characters of
classical groups 
(see \cite{Tsuchiya,Br,Ku1,Ku2,Okada,RS,CP,BWZ,BFK} for examples).
The seminal works are by Tsuchiya \cite{Tsuchiya} and Kuperberg \cite{Ku1,Ku2},
in which they found determinant and Pfaffian representations
for various variations of the DWBPF.
There are also works on the free-fermionic model
\cite{Okadareflecting,HK1,HK2,ZZfelderhof,FCWZfelderhof,FWZfelderhof,Zuparicfelderhof,Wheelerfelderhof,BSfelderhof},
in which more simplified factorized representations of the partition functions were found, even for elliptic models.

Studying elliptic generalizations of the DWBPF is interesting,
and it is particularly interesting to find determinant and Pfaffian representations.
In particular, finding representations using Pfaffians of a matrix
whose matrix entries are elliptic functions is interesting,
since there are only a few studies on elliptic Pfaffians.
For example, Rosengren \cite{Rosell1} introduced a family of
elliptic Pfaffians
and showed that the partition functions of the Andrews-Baxter-Forrester (ABF) model \cite{ABFfelderhof} at the supersymmetric point
are expressed as a sum of two elliptic Pfaffians.
We mention that expressions of the DWBPF of the ABF model which hold in generic parameters
are derived in \cite{Ros,PRS,YZ}, a factorized expression at the free-fermion
point
is derived in \cite{FWZfelderhof,Zuparicfelderhof,Wheelerfelderhof}.
and a single determinant representation was recently derived in
\cite{Galleasthree}.

As for the properties of elliptic Pfaffians,
Okada \cite{Okadapfaffian}, Rosengren
\cite{Roselldet,Rosellpfaffian} and Rains \cite{Rains}
discovered several elliptic generalizations of the Pfaffian counterpart
\cite{Schur} of the Cauchy determinant formulas.
The properties of elliptic determinants have been extensively studied.
For example, several
generalizations of the Cauchy determinant formula \cite{Cauchy}
have been discovered \cite{Frob,Has,War,RosSch}.
On the other hand, there are only a few results on elliptic Pfaffians
by Okada, Rosengren and Rains.

In this paper, we study partition functions of an elliptic free-fermionic
face model with a triangular boundary,
and show that they can be explicitly expressed using elliptic Pfaffians.
The face model we treat can be regarded as degenerations of
the ABF model \cite{ABFfelderhof},
Okado-Deguchi-Martin (elliptic Perk-Schultz) model \cite{Okado,DMfelderhof}
and
Foda-Wheeler-Zuparic (elliptic Felderhof) model \cite{FWZfelderhof},
which are face-type counterparts of the elliptic vertex models
\cite{eightvertex,Felderhoffelderhof},
and are elliptic analogues of the trigonometric models
of the $U_q(sl_2)$ six-vertex model,
Perk-Schultz model and the Felderhof free-fermion model
\cite{Dr,J,LW,PSfelderhof,Murakami,DAfelderhof}.
In this paper, we treat a fundamental example of the variations of the DWBPF
introduced by Kuperberg \cite{Ku2}.
Kuperberg introduced a class of partition functions of the $U_q(sl_2)$
six-vertex model with a triangular boundary
using an off-diagonal boundary $K$-matrix at the boundary,
and showed that they have explicit expressions using Pfaffians.
He called this boundary condition the OS boundary.
We introduce the partition functions of the
elliptic free-fermionic face model with OS boundary,
and study them using the elliptic version of the Izergin-Korepin analysis.
We evaluate the explicit representations of the
partition functions using elliptic Pfaffians,
and we get two Pfaffian representations based on two versions
of the Izergin-Korepin analysis.
The Izergin-Korepin analysis for various types of
partition functions of trigonometric models
\cite{Ko,Iz,Tsuchiya,Ku1,Ku2,Wheeler}
and a closely related functional equation approach
have been extended to elliptic models, and have been used to compute the DWBPF,
wavefunctions and scalar products
of elliptic integrable models \cite{Ros,PRS,YZ,FK,Filali,Chinesegroup,Chinesegroup2,Galleasone,Galleastwo,GL,Lamers,Moellipticfelderhof,MotegiIzerginKorepin,Motegiscalarproducts} in recent years.
As a corollary of
the two elliptic Pfaffian representations of the same partition functions
by the elliptic Izergin-Korepin analysis,
we get an identity between the two elliptic Pfaffians.

This paper is organized as follows.
In the next section, we introduce and summarize formulas and properties
of the Pfaffian and theta functions which will be used in later sections.
In section 3, we introduce the elliptic free-fermionic face model
using the dynamical $R$-matrix formalism, and
introduce the partition functions with OS boundary.
In section 4, we analyze and get the explicit expressions
of the partition functions using elliptic Pfaffians.
We also get an identity between the two elliptic Pfaffians
as a corollary of the two representations of the partition functions.
Section 5 is devoted to the conclusion of this paper.

\section{Preliminaries}
In this section, 
we introduce and present some formulas and properties
of the Pfaffian and theta functions,
which are going to be used
in this paper for the analysis of the DWBPF with OS boundary.

The Pfaffian $\mathrm{Pf} X$ of a skew-symmetric matrix
$X=(x_{ij})_{1 \le i,j \le 2n}$ is defined as
\begin{align}
\mathrm{Pf} X=\sum_{\sigma \in M_{2n}}
\mathrm{sgn}(\sigma) \prod_{j=1}^n x_{\sigma(2j-1) \ \sigma(2j)},
\label{definitionpfaffian}
\end{align}
where $M_{2n}$ is a subset of the symmetric group $S_{2n}$
satisfying
\begin{align}
M_{2n}=\Bigg\{ \sigma \in S_{2n} \ \Bigg| \
\sigma(1) < \sigma(3) < \cdots < \sigma(2n-1),
\ \sigma(2j-1) < \sigma(2j), \ j=1,\dots,n \Bigg\}.
\end{align}
In this paper, besides the definition of the Pfaffian,
we use the following expansion formula for the Pfaffian
\begin{align}
\mathrm{Pf}X=\sum_{k=2}^{2n} (-1)^k x_{1k}
\mathrm{Pf}X_{1,k}^{1,k}. \label{pfaffianexpansion}
\end{align}
Here $X_{1,k}^{1,k}$ is a $(2n-2) \times (2n-2)$ matrix
in which the first and $k$-th rows and columns are removed from the
$2n \times 2n$ matrix $X=(x_{ij})_{1 \le i,j \le 2n}$.

We introduce the notation $[u]$ for the theta functions $[u]=H(\pi i u)$
where $H(u)$ is
\begin{align}
H(u)=2 \sinh u
\prod_{j=1}^\infty (1-2{\bf q}^{2j} \cosh 2u+{\bf q}^{4j})(1-{\bf q}^{2j}).
\end{align}
Here, ${\bf q}$ is the elliptic nome $(0 < {\bf q} < 1)$.

The theta function $[u]$ is an odd function $[-u]=-[u]$
and hence $[0]=0$.
It also satisfies the quasi-periodicities
\begin{align}
[u+1]&=-[u], \label{qpone} \\
[u-i \log ({\bf q})/\pi]&=-{\bf q}^{-1}
\exp  (-2 \pi i u) [u] \label{qptwo}.
\end{align}
Using the above properties, we get
\begin{align}
[u-1/2]=[-u-1/2], \label{usethisproperty}
\end{align}
for example.
The addition formula for the theta functions
\begin{align}
[u+x][u-x][v+y][v-y]-[v+x][v-x][u+y][u-y]-[x+y][x-y][u+v][u-v]=0,
\label{additionformula}
\end{align}
is one of the most important identities for the theta functions.
For example, it is used to prove the Yang-Baxter relation
for elliptic integrable models.

The following facts about the elliptic polynomials \cite{PRS,FSfelderhof}
turned out to be useful for the analysis of
elliptic face-type integrable models \cite{ABFfelderhof}.
They were used in developing the method of quantum separation of variables
for the ABF model and the elliptic Gaudin model \cite{FSfelderhof}.
These facts justify the Izergin-Korepin analysis on elliptic integrable models
and were used effectively on the computation of the DWBPF of
elliptic integrable models. See Refs.~\cite{PRS}, \cite{Ros}, and \cite{YZ} for examples.

A character is a group homomorphism
$\chi$ from multiplicative groups
$\Gamma=\mathbf{Z}+\tau \mathbf{Z}$ to $\mathbf{C}^\times$.
For each character $\chi$ and positive integer $n$, an $n$-dimensional space $\Theta_n(\chi)$
is defined 
that consists of holomorphic functions $\phi(y)$ on $\mathbf{C}$
satisfying the quasiperiodicities
\begin{align}
\phi(y+1)&=\chi(1) \phi(y), \label{propertyuseone} \\
\phi(y+\tau)&=\chi(\tau) e^{-2 \pi i ny-\pi i n \tau}\phi(y).
\label{propertyusetwo}
\end{align}
The elements of the space $\Theta_n(\chi)$ are called elliptic polynomials.
The space $\Theta_n(\chi)$ is $n$-dimensional \cite{PRS,FSfelderhof},
and the following fact holds for the elliptic polynomials:
\begin{proposition} \cite{PRS,FSfelderhof} \label{propositionelliptic}
Suppose there are two elliptic polynomials $P(y)$ and $Q(y)$
in $\Theta_n(\chi)$, where $\chi(1)=(-1)^n$ and $\chi(\tau)=(-1)^n e^\alpha$.
If these two polynomials are equal at $n$ points $y_j$, $j=1,\dots,n$, satisfying
$y_j-y_k \not\in \Gamma$ and $\sum_{k=1}^N y_k-\alpha \not\in \Gamma$, that is, $P(y_j)=Q(y_j)$,
then the two polynomials are exactly the same: $P(y)=Q(y)$.
\end{proposition}

\section{Elliptic free-fermionic face model}\label{sec2}
In this section, we introduce the free-fermionic face model
using the dynamical $R$-matrix formalism
\cite{Felderfelderhof,FelderVarchenko,FWZfelderhof,Zuparicfelderhof,Wheelerfelderhof},
which enables one to describe the face model like a six-vertex model.

\begin{figure}[ht]
\includegraphics[width=13.5cm]{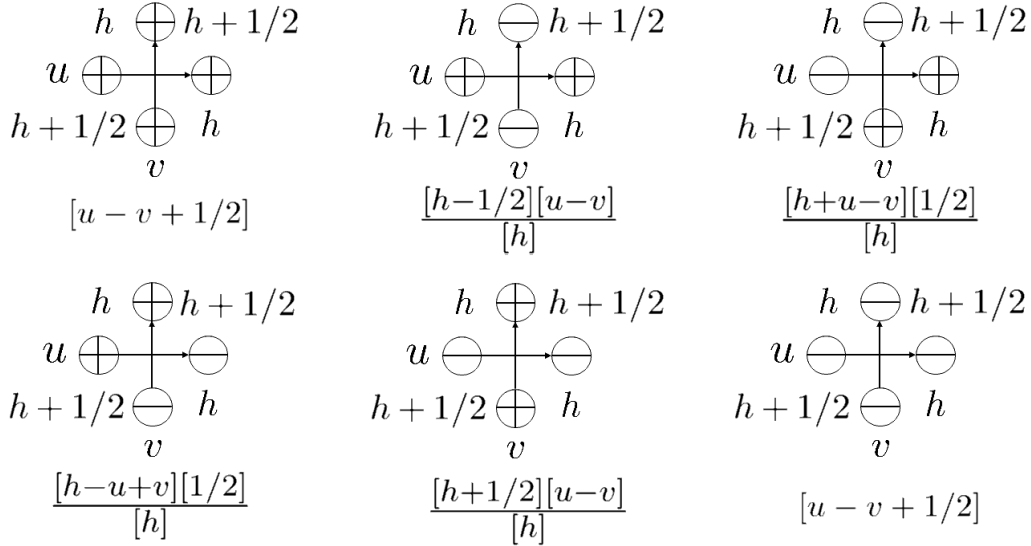}
\caption{The free-fermionic dynamical $R$-matrix $R_{ab}(u,v|h)$, \eqref{rmatrix}.
Each line is to be regarded as a representation space
and carries a spectral parameter.
In this picture, the horizontal lines carry
a spectral parameter $u$, while the vertical lines
carry $v$. The height variables of neighboring regions (regions separated by a line) differ by 1/2. For the case of the free-fermion model,
we can think that the regions take height variables either $h$ or $h+1/2$,
due to the property of the dynamical $R$-matrix $R(u,v|h+1)=R(u,v|h)$.
The (dual) basis vector $|0\rangle$ ($\langle 0|$)
is depicted as $\oplus$, while $|1\rangle$ ($\langle 1|$)
is depicted as $\ominus$.}
\label{picturegeneralizedloperator}
\end{figure}

The dynamical $R$-matrix of the elliptic free-fermionic face model
is given by (see Fig. \ref{picturegeneralizedloperator})
\begin{align}
R_{ab}(u,v|h)
=\begin{pmatrix}
[u-v+1/2] & 0 & 0 & 0 \\
0 & \frac{[h-1/2][u-v]}{[h]} &
\frac{[h+u-v][1/2]}{[h]}  & 0 \\
0 & \frac{[h-u+v][1/2]}{[h]} & 
\frac{[h+1/2][u-v]}{[h]} & 0 \\
0 & 0 & 0 & [u-v+1/2]
\end{pmatrix},
\label{rmatrix}
\end{align}
acting on the tensor product $W_a \otimes W_b$
of the complex two-dimensional space $W_a$.
The free-fermionic dynamical $R$-matrix satisfies
$R_{ab}(u,v|h+1)=R_{ab}(u,v|h)$.

\begin{figure}[ht]
\includegraphics[width=13.5cm]{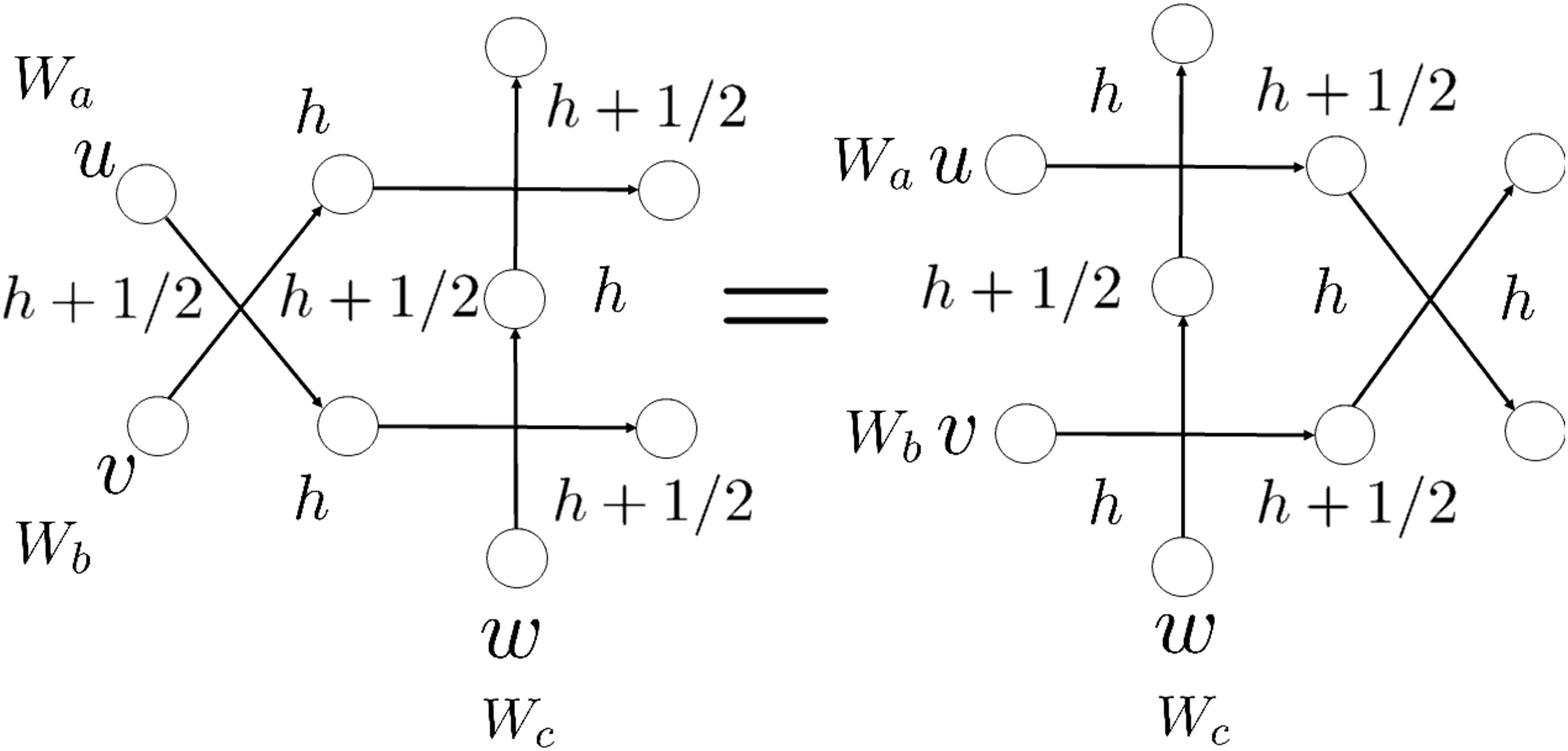}
\caption{The free-fermionic dynamical Yang--Baxter relation \eqref{yangbaxter}.
The left- and right-hand sides of the figure represent
the left- and right-hand sides of the Yang-Baxter relation
$R_{bc}(v,w|h)R_{ac}(u,w|h+1/2)R_{ab}(u,v|h)$
and
$R_{ab}(u,v|h+1/2)R_{ac}(u,w|h)R_{bc}(v,w|h+1/2)$,
respectively.
}
\label{pictureyangbaxter}
\end{figure}

The dynamical $R$-matrix
\eqref{rmatrix}
satisfies the dynamical
Yang--Baxter relation (Fig.~\ref{pictureyangbaxter})
\begin{align}
&R_{bc}(v,w|h)R_{ac}(u,w|h+1/2)R_{ab}(u,v|h)  \\
=&R_{ab}(u,v|h+1/2)R_{ac}(u,w|h)R_{bc}(v,w|h+1/2),
\label{yangbaxter}
\end{align}
acting on $W_a \otimes W_b \otimes W_c$.

\begin{figure}[ht]
\includegraphics[width=10cm]{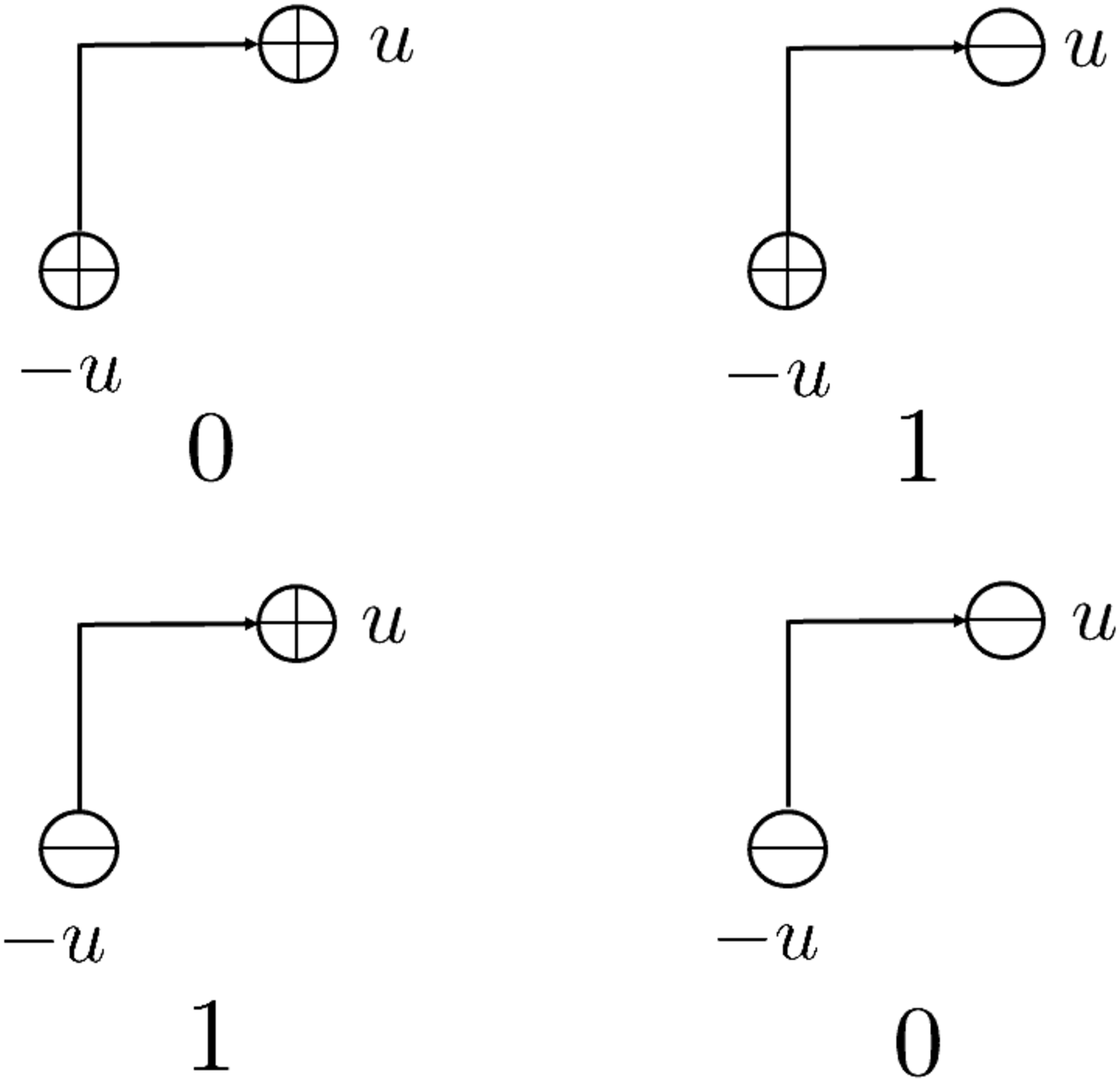}
\caption{The off-diagonal $K$-matrix $K(u,h)$, \eqref{kmatrix}.
The horizontal lines carry a spectral parameter $u$,
while the vertical lines carry $-u$.
}
\label{picturekmatrix}
\end{figure}

We also introduce the following off-diagonal $K$-matrix acting on
$W_a$ (see Fig.~\ref{picturekmatrix}):
\begin{align}
K_{a}(u,h)=\begin{pmatrix}
0 & 1 \\
1 & 0 \\
\end{pmatrix}, \label{kmatrix}
\end{align}

\begin{figure}[ht]
\includegraphics[width=10cm]{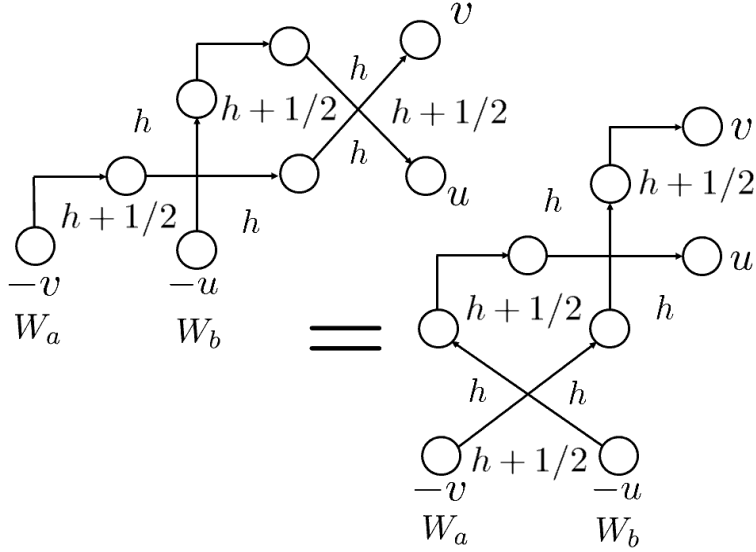}
\caption{
The reflection equation \eqref{reflection equation}.
The left- and right-hand sides of the figure represent
the left- and right-hand sides of the reflection equation
$R_{ba}(u-v,h)K_b(u,h)R_{ab}(v+u,h)K_a(v,h)$ and
$K_a(v,h)R_{ba}(u+v,h)K_b(u,h)R_{ab}(u-v,h)$,
respectively.
}
\label{picturereflection}
\end{figure}

One can easily check that the
$K$-matrix \eqref{kmatrix} together with the dynamical
$R$-matrix \eqref{rmatrix}
satisfy the relation
\begin{align}
R_{ba}(u-v,h)K_b(u,h)R_{ab}(v+u,h)K_a(v,h)
=K_a(v,h)R_{ba}(u+v,h)K_b(u,h)R_{ab}(u-v,h),
\label{reflection equation}
\end{align}
which is called the reflection equation
or the boundary Yang--Baxter equation \cite{Sklyanin}
(Fig.~\ref{picturereflection}).
The reflection equation ensures integrability at the boundary.
This off-diagonal $K$-matrix was used
as local pieces of the partition functions
for the case of the $U_q(sl_2)$ six-vertex model by Kuperberg \cite{Ku2}.
We also use this $K$-matrix for the elliptic integrable model
in this paper.
It seems that it is hard or maybe
impossible to extract the off-diagonal $K$-matrix
\eqref{kmatrix}
from the general full $K$-matrices of elliptic integrable models
\cite{VGR,IK,FHS,BP}.
This $K$-matrix was used to impose the antiperiodic boundary condition
on the ABF model in the paper by Felder-Schorr \cite{FSfelderhof,Schorr},
in which they analyzed the antiperiodic boundary condition
by the quantum separation of variables method.

\begin{figure}[ht]
\includegraphics[width=10cm]{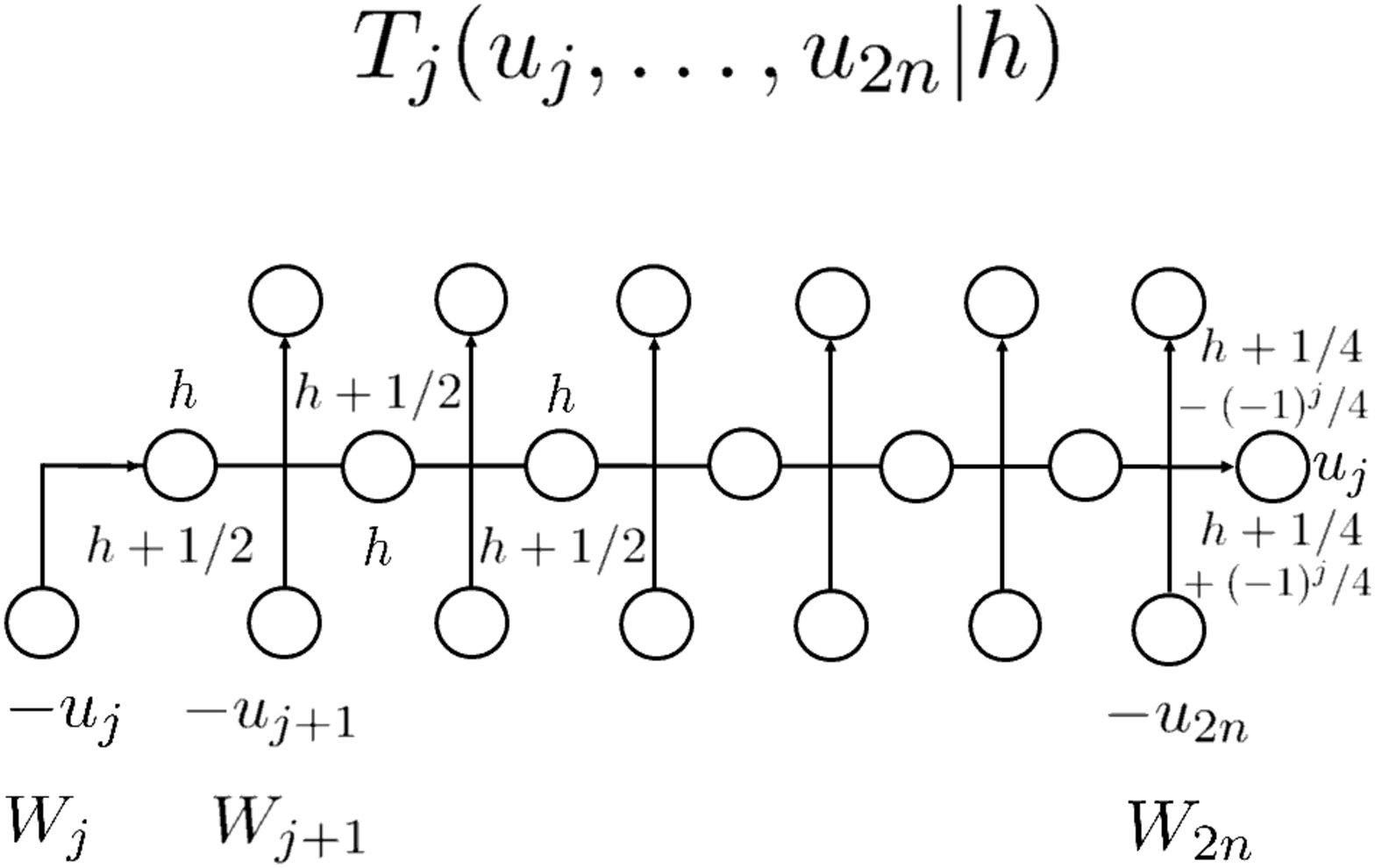}
\caption{The monodromy matrix $T_j(u_j,\dots,u_{2n}|h)$,
\eqref{monodromy1},
constructed from one
$K$-matrix \eqref{kmatrix} and $2n-j$ dynamical
$R$-matrices \eqref{rmatrix}
.}
\label{picturemonodromy}
\end{figure}

\section{Partition functions with OS boundary}\label{sec3}

In this section, we introduce and analyze the partition functions
of the free-fermionic face model with OS boundary.

Let us denote the orthonormal basis of $W_a$ and its dual by
$\{|0 \rangle_a, |1 \rangle_a \}$ and $\{{}_a \langle 0|, {}_a \langle 1|\}$.
Next, the  Pauli spin operators
$\sigma^+$ and $\sigma^-$ are defined as operators acting on the (dual) orthonormal
basis as
\begin{align}
\sigma^+|1 \rangle&=|0 \rangle, & 
\sigma^+|0 \rangle&=0, & 
\langle 0|\sigma^+&=\langle 1|, &
\langle 1|\sigma^+&=0, 
\\
\sigma^-|0 \rangle&=|1 \rangle,&
\sigma^-|1 \rangle&=0, &
\langle 1|\sigma^-&=\langle 0|,&
\langle 0|\sigma^-&=0.
\end{align}

To formulate the wavefunctions with a
triangular boundary, we introduce the tensor product of the Fock spaces:
$W_{1} \otimes \cdots \otimes W_{2n}$.

\begin{figure}[ht]
\includegraphics[width=11cm]{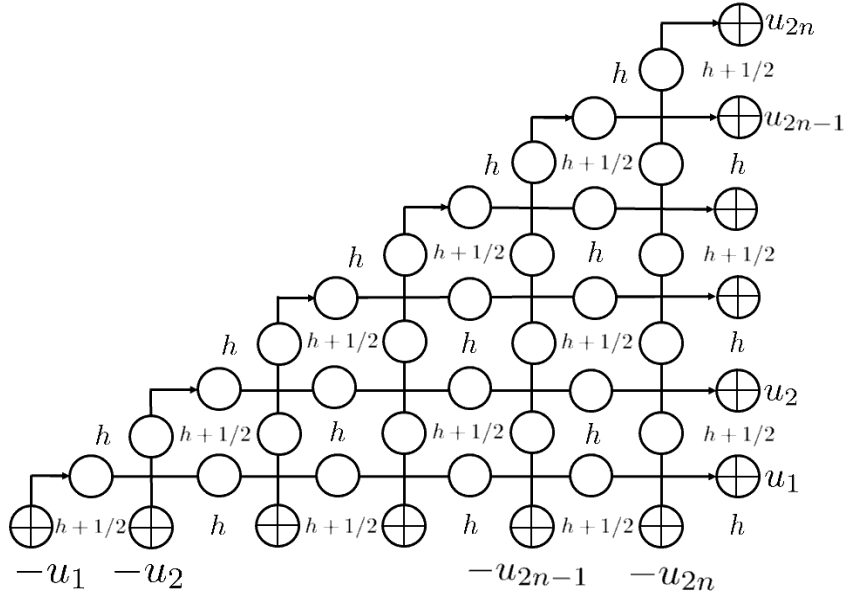}
\caption{Partition functions
$P_{2n}(u_1,\dots,u_{2n}|h)$,
\eqref{partitionfunctions}, with OS boundary.
}
\label{OSboundary}
\end{figure}

Using the dynamical $R$-matrix \eqref{rmatrix} and the
$K$-matrix \eqref{kmatrix}, we  next define a monodromy matrix
$T_j(u_j,\dots,u_{2n}|h)$, $j=1,\dots,2n$, as 
\begin{align}
T_{j}(u_j,\dots,u_{2n}|h)
=
\prod_{k=j+1}^{2n} R_{jk}(u_j,-u_{k}|h+1/4+(-1)^{k-j}/4)
K_{j}(u_j),
\label{monodromy1}
\end{align}
which acts on $W_{j} \otimes \cdots \otimes W_{2n}$.
See Fig.~\ref{picturemonodromy}
for a pictorial depiction of \eqref{monodromy1}.
Using this monodromy matrix,
we introduce the partition functions
$P_{2n}(u_1,\dots,u_{2n}|h)$ as follows
(Fig. \ref{OSboundary})
:
\begin{align}
P_{2n}(u_1,\dots,u_{2n},h)
={}_{2n} \langle \Omega|
T_{2n}(u_{2n}|h) \cdots T_1(u_1,\dots,u_{2n}|h)
|\Omega \rangle_{2n},
\label{partitionfunctions}
\end{align}
where the states ${}_{2n} \langle \Omega|$ and $|\Omega \rangle_{2n}$
are defined as
\begin{align}
{}_{2n} \langle \Omega|={}_{1} \langle 0| \otimes \cdots \otimes
{}_{2n} \langle 0|,
\\
|\Omega \rangle_{2n}=|0 \rangle_{1} \otimes \cdots \otimes
|0 \rangle_{2n}.
\end{align}

\begin{figure}[ht]
\includegraphics[width=11cm]{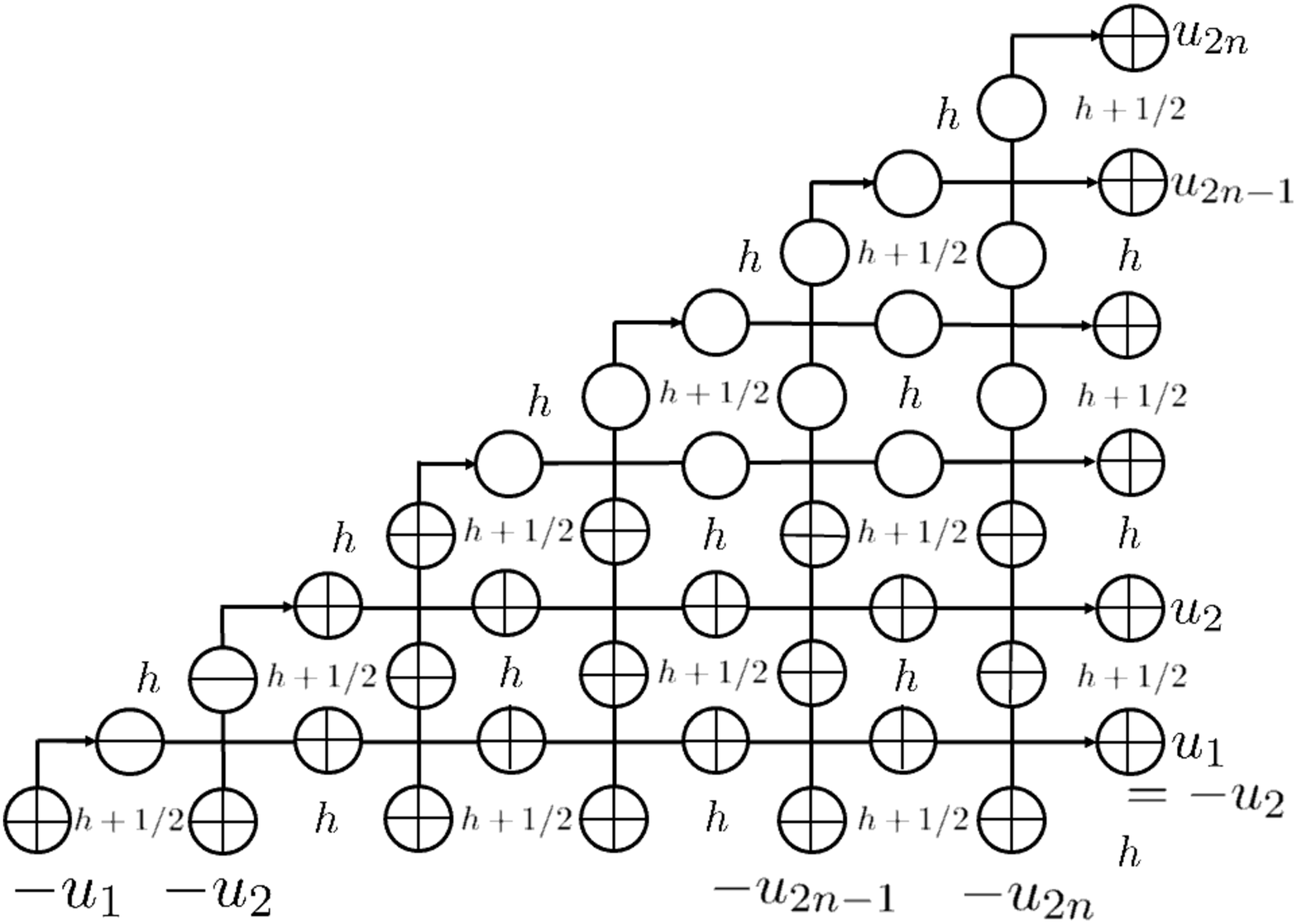}
\caption{Partition functions
$P_{2n}(u_1,\dots,u_{2n}|h)$
evaluated at $u_1=-u_2$
\eqref{ordinaryrecursionwavefunction}.
The bottom two rows are frozen due to the
properties of the dynamical $R$-matrix \eqref{rmatrix}
and the $K$-matrix \eqref{kmatrix}.
}
\label{picturerecursion}
\end{figure}

\begin{figure}[ht]
\includegraphics[width=11cm]{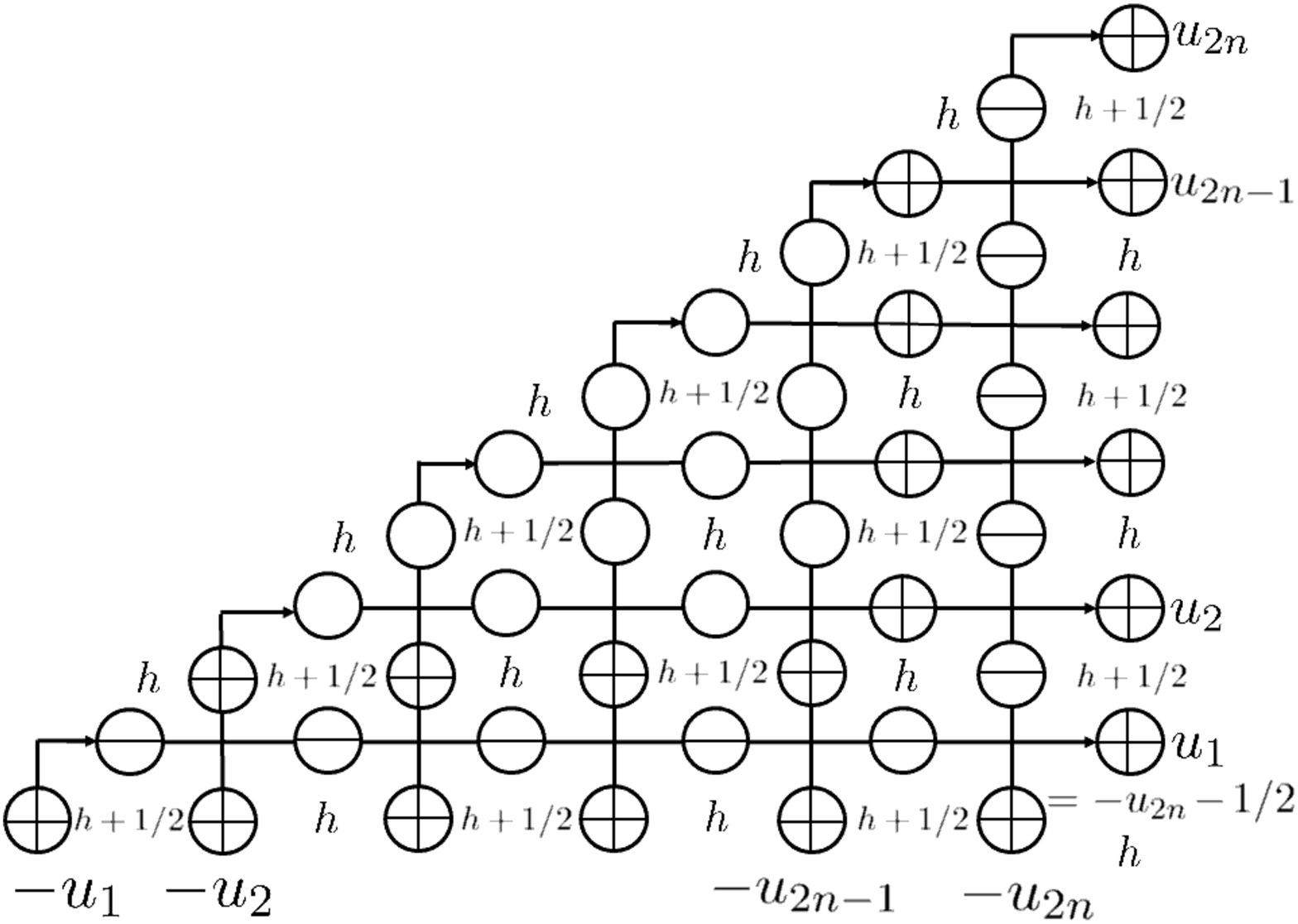}
\caption{Partition functions
$P_{2n}(u_1,\dots,u_{2n}|h)$
evaluated at $u_1=-u_{2n}-1/2$
\eqref{anotherrecursion}.
The bottom row and the rightmost column are frozen due to the
properties of the dynamical $R$-matrix \eqref{rmatrix}
and the $K$-matrix \eqref{kmatrix}.
}
\label{picturerecursiontwo}
\end{figure}

Now we perform the Izergin--Korepin analysis \cite{Ko,Iz} on
the partition functions \\
$P_{2n}(u_1,\dots,u_{2n}|h)$.
The Izergin-Korepin analysis is a technique introduced by
Korepin, and the idea is to list the properties of the domain wall boundary partition functions which uniquely determine them,
and reduce the problem of explicitly computing the partition functions
to that of finding the polynomials satisfying those properties.
The Izergin-Korepin technique needs the notion of degree of the polynomial
for the uniqueness,
and the notion of the degree and the property of the elliptic polynomial
stated in Proposition \ref{propositionelliptic}
in Section 2 ensures
the elliptic version of the Izergin-Korepin analysis,
which was effectively used for the computation of the
ordinary domain wall boundary partition functions
of the Andrews-Baxter-Forrester model \cite{Ros,PRS,YZ}.

\begin{proposition} 
\label{ordinarypropertiesfordomainwallboundarypartitionfunction}
The partition functions with OS boundary
$P_{2n}(u_1,\dots,u_{2n}|h)$
satisfy, and are uniquely determined by, the following properties:
\begin{enumerate}
\item The partition functions
$P_{2n}(u_1,\dots,u_{2n}|h)$
are elliptic polynomials in $u_{1}$ of degree $2n-1$ with
the following quasi-periodicities: 
\begin{align}
&P_{2n}(u_1+1,\dots,u_{2n}|h)
=(-1)^{2n-1} P_{2n}(u_1,\dots,u_{2n}|h), \label{qppartitionfunctionsone} \\
&P_{2n}(u_1-i \log ({\bf q})/\pi,\dots,u_{2n}|h)
 \nonumber\\
={}&(-{\bf q}^{-1})^{2n-1}
\exp \! \Bigg(-2 \pi i
\Bigg((2n-1)u_1+h+\sum_{j=2}^{2n} u_j
\Bigg)
\Bigg) P_{2n}(u_1,\dots,u_{2n}|h) \label{qppartitionfunctionstwo}.
\end{align}
\item The following relations among the
partition functions hold (Fig. \ref{picturerecursion}):
\begin{align}
&\hspace{-36pt}P_{2n}(u_1,\dots,u_{2n}|h)
|_{u_1=-u_\ell}
\nonumber \\
={}&[1/2]
\prod_{ \substack{ j=2 \\ j \neq \ell} }^{2n}[u_j+u_\ell+1/2][u_j-u_\ell+1/2]
P_{2n-2}(u_2,\dots,\hat{u_\ell} ,\dots,u_{2n}|h)
, \label{ordinaryrecursionwavefunction}
\end{align}
for $\ell=2,\dots,2n$,
and $\hat{u_\ell}$ in $P_{2n-2}(u_2,\dots,\hat{u_\ell} ,\dots,u_{2n}|h)$
means that $\hat{u_\ell}$ is removed.
\item  The following evaluation holds:
\begin{align}
&\hspace{-36pt}P_{2}(u_1,u_2|h)=
\frac{[1/2][h+u_1+u_2]}{[h]}.
\label{ordinaryinitialrecursion}
\end{align}
\end{enumerate}
\end{proposition}

\begin{proof}
Before going into the details of proving Properties (1)--(3),
let us first point out that they uniquely determine the partition functions $P_{2n}(u_1,\dots,u_{2n}|h)$.
The uniqueness follows by using induction on $n$, together with
the recurrence of Property (2) and the initial condition of Property (3).
Property (2) connects $2n-1$ special points $u_1=-u_\ell$, $\ell=2,\dots,2n$
of $P_{2n}(u_1,\dots,u_{2n}|h)$ with
$P_{2n-2}(u_2,\dots,\hat{u_\ell},\dots,u_{2n}|h)$.
These evaluations at $2n-1$ special points and
Property (1) together with Proposition \ref{propositionelliptic} imply that
$P_{2n}(u_1,\dots,u_{2n}|h)$ is uniquely determined from
$P_{2n-2}(u_2,\dots,\hat{u_\ell},\dots,u_{2n}|h)$ by Properties (1) and (2).
Property (3) corresponds to the initial term of the recurrence relation.

Now let us go into the details of the proof.
Properties (1)--(3)
can be proved in the standard way.

We first show Property (1).
We use the completeness relation
in one up-spin sector,
\begin{align}
\sum_{j=1}^{2n-1}
|0^{j-1}10^{2n-j-1} \rangle \langle 0^{j-1}10^{2n-j-1} |=\mathrm{Id}, 
\end{align}
on the space $W_2 \otimes \cdots \otimes W_{2n}$
with
\begin{align*}
|0^{j-1}10^{2n-j-1} \rangle&=|0 \rangle_{2} \otimes \cdots \otimes
|0 \rangle_{j} \otimes |1 \rangle_{j+1} \otimes
|0 \rangle_{j+2} \otimes \cdots \otimes |0 \rangle_{2n}, \\
\langle 0^{j-1}10^{2n-j-1} |&={}_{2} \langle 0 | \otimes \cdots \otimes
{}_{j} \langle 0 | \otimes {}_{j+1} \langle 1 | \otimes
{}_{j+2} \langle 0 | \otimes \cdots \otimes {}_{2n} \langle 0 |,
\end{align*}
and decompose $P_{2n}(u_1,\dots,u_{2n}|h)$ as 
\begin{align}
&\hspace{-24pt}P_{2n}(u_1,\dots,u_{2n}|h) \nonumber \\
={}&\sum_{j=1}^{2n-1}
{}_{2n-1} \langle \Omega|
T_{2n}(u_{2n}|h) \cdots 
T_2(u_2,\dots,u_{2n}|h)
|0^{j-1}10^{2n-j-1} \rangle \nonumber \\
&\times {}_{1} \langle 0| \otimes
\langle 0^{j-1}10^{2n-j-1}|
T_1(u_1,\dots,u_{2n}|h)
|\Omega \rangle_{2n},
\label{anotherexpressionexpansion}
\end{align}
where ${}_{2n-1} \langle \Omega|={}_{2} \langle 0| \otimes \cdots \otimes
{}_{2n} \langle 0|$.

One can easily calculate the explicit forms of
${}_{1} \langle 0| \otimes
\langle 0^{j-1}10^{2n-j-1}|
T_1(u_1,\dots,u_{2n}|h)
|\Omega \rangle_{2n}:=f_j(u_1)$ in
\eqref{anotherexpressionexpansion},
which are
\begin{align}
f_j(u_1)
={}&\frac{[1/2][h+(j-1)/2+u_1+u_{j+1}]}{[h]}
\nonumber \\
&\times \prod_{k=2}^{j} [u_1+u_k]
\prod_{k=j+2}^{2n} [u_1+u_k+1/2]. \label{explicitforqp}
\end{align}
It is easy to see from
\eqref{explicitforqp} and
the quasi-periodicities of the theta functions
\eqref{qpone} and \eqref{qptwo} that the quasi-periodicities
of $f_j(u_1)$ are
\begin{align}
&f_j(u_1+1) 
=(-1)^{2n-1} f_j(u_1), \label{qpfirst}\\
&f_j(u_1-i \log ({\bf q})/\pi)
 \nonumber\\
={}&(-{\bf q}^{-1})^{2n-1}
\exp \! \Bigg(-2 \pi i
\Bigg((2n-1)u_1+h+\sum_{\ell=2}^{2n} u_{\ell}
\Bigg)
\Bigg) f_j(u_1). \label{qpsecond}
\end{align}
Since the quasi-periodicities for $f_j(u_1)$,
\eqref{qpfirst}, and \eqref{qpsecond}, do not depend on $j$,
and noting that the dependence on $u_1$ for each summand in
the right hand side of \eqref{anotherexpressionexpansion},
comes only from $f_j(u_1)$,
one finds that the
quasi-periodicities of the partition functions $P_{2n}(u_1,\dots,u_{2n}|h)$
are given by \eqref{qppartitionfunctionsone} and
\eqref{qppartitionfunctionstwo}.
One also concludes from \eqref{qppartitionfunctionsone} and
\eqref{qppartitionfunctionstwo} that
$P_{2n}(u_1,\dots,u_{2n}|h)$
are elliptic polynomials in $u_1$ of degree $2n-1$.

Next, we prove property~(2).
First, one shows
\eqref{ordinaryrecursionwavefunction} for the case $\ell=2$
by using a graphical representation
of the partition functions (Fig. \ref{picturerecursion}),
as is always the case when using the Korepin's method.
First, one observes that the $K$-matrix at the bottom row
is already frozen since the $K$-matrix we use 
for the partition fuctions $P_{2n}(u_1,\dots,u_{2n}|h)$
is an off-diagonal one \eqref{kmatrix}.
If we set $u_1=-u_2$, one finds that the $R$-matrix adjacent to the frozen
$K$-matrix gets frozen since
${}_1 \langle 1 | {}_2 \langle 0 | R_{12}(-u_2,-u_2,h)|1 \rangle_1 | 0 \rangle_2=0$, and continuing graphical observation
using the ice-rule
\begin{align}
{}_a \langle \gamma | {}_b \langle \delta | R_{ab}(u,v,h)|\alpha \rangle_a | \beta \rangle_b=0, \qquad
\text{unless}\quad\alpha+\beta=\gamma+\delta, \label{icerule}
\end{align}
one sees that the two bottom rows freeze (Fig. \ref{picturerecursion}).
The product of the matrix elements of the $R$-matrices
of the frozen two rows is $\displaystyle [1/2]
\prod_{ j=3 }^{2n}[u_j+u_2+1/2][u_j-u_2+1/2]$,
and the remaining unfrozen part is
$P_{2n-2}(u_3,\dots,u_{2n}|h)$, and we get
\begin{align}
&\hspace{-36pt}P_{2n}(u_1,\dots,u_{2n}|h)
|_{u_1=-u_2}
\nonumber \\
={}&[1/2]
\prod_{ j=3 }^{2n}[u_j+u_2+1/2][u_j-u_2+1/2]
P_{2n-2}(u_3,\dots,u_{2n}|h). \label{basecase}
\end{align}
One can show that the partition functions
$P_{2n}(u_1,\dots,u_{2n}|h)$
are symmetric with respect to the spectral parameters
$u_1,\dots,u_{2n}$
by the standard railroad argument using
the dynamical Yang-Baxter relation and the
reflection equation (Kuperberg \cite{Ku2},
see also \cite{MotegiIzerginKorepin}).
From \eqref{basecase} and the symmetry property,
one finds that \eqref{ordinaryrecursionwavefunction} holds.

Finally, it is trivial to check Property (3) from the definition
of the $R$-matrix \eqref{rmatrix}.

\end{proof}
One can prove that there are explicit expressions for the
partition functions with OS boundary $P_{2n}(u_1,\dots,u_{2n}|h)$
in terms of elliptic Pfaffians
by showing that the right hand side of \eqref{ellipticrepresentation}
satisfies all the properties in
Proposition~\ref{ordinarypropertiesfordomainwallboundarypartitionfunction}.

\begin{theorem} \label{maintheorem}
The partition functions with OS boundary $P_{2n}(u_1,\dots,u_{2n}|h)$
have the following expressions
using elliptic Pfaffians:
\begin{align}
&P_{2n}(u_1,\dots,u_{2n}|h) \nonumber \\
=&\prod_{1 \le i < j \le 2n}
\frac{[u_j+u_i][u_j-u_i+1/2]}{[u_j-u_i]}
\mathrm{Pf} \Bigg( 
\frac{[1/2][u_j-u_i][u_i+u_j+h]}{[h][u_i+u_j][u_j-u_i+1/2]}
\Bigg)_{1 \le i,j \le 2n}. \label{ellipticrepresentation}
\end{align}
\end{theorem}
Note that 
$\displaystyle \Bigg( 
\frac{[1/2][u_j-u_i][u_i+u_j+h]}{[h][u_i+u_j][u_j-u_i+1/2]}
\Bigg)_{1 \le i,j \le 2n}$ is a skew-symmetric matrix
which can be checked using the facts that $[u]$ is an odd function
and the property \eqref{usethisproperty}.

\begin{proof}
Let us denote the right hand side of \eqref{ellipticrepresentation}
as $E_{2n}(u_1,\dots,u_{2n}|h)$:
\begin{align}
&E_{2n}(u_1,\dots,u_{2n}|h) \nonumber \\
:=&\prod_{1 \le i < j \le 2n}
\frac{[u_j+u_i][u_j-u_i+1/2]}{[u_j-u_i]}
\mathrm{Pf} \Bigg( 
\frac{[1/2][u_j-u_i][u_i+u_j+h]}{[h][u_i+u_j][u_j-u_i+1/2]}
\Bigg)_{1 \le i,j \le 2n}. \label{definitionofE}
\end{align}
We show that $E_{2n}(u_1,\dots,u_{2n}|h)$
satisfies all the properties in
Proposition~\ref{ordinarypropertiesfordomainwallboundarypartitionfunction}.
Let us show Property (1).
To check this, we view $E_{2n}(u_1,\dots,u_{2n}|h)$ as
a function of $u_1$ and split the function as
$E_{2n}(u_1,\dots,u_{2n}|h)=e_1(u_1)e_2(u_1)$,
where $e_1(u_1)$ and $e_2(u_1)$ are
the overall factor and the elliptic Pfaffian, respectively
\begin{align}
e_1(u_1)&=\prod_{1 \le i < j \le 2n}
\frac{[u_j+u_i][u_j-u_i+1/2]}{[u_j-u_i]}, \\
e_2(u_1)&=\mathrm{Pf} \Bigg( 
\frac{[1/2][u_j-u_i][u_i+u_j+h]}{[h][u_i+u_j][u_j-u_i+1/2]}
\Bigg)_{1 \le i,j \le 2n}.
\end{align}

Using the quasi-periodicities of the theta functions
\eqref{qpone} and \eqref{qptwo},
it is easy to calculate
the quasi-periodicites of the overall factor $e_1(u_1)$:
\begin{align}
&e_{1}(u_1+1)
=(-1)^{2n-1} e_{1}(u_1), \label{qpcheckone} \\
&e_{1}(u_1-i \log ({\bf q})/\pi)
 \nonumber\\
={}&(-{\bf q}^{-1})^{2n-1}
\exp \! \Bigg(-2 \pi i
\Bigg((2n-1)u_1+\sum_{j=2}^{2n} u_j-1/2
\Bigg)
\Bigg) e_{1}(u_1). \label{qpchecktwo}
\end{align}
Next, noting that the quasi-periodicities of
the matrix elements of the first row of the matrix
$\displaystyle X=\Bigg( 
\frac{[1/2][u_j-u_i][u_i+u_j+h]}{[h][u_i+u_j][u_j-u_i+1/2]}
\Bigg)_{1 \le i,j \le 2n}$
are given by
\begin{align}
&\frac{[1/2][u_j-(u_1+1)][(u_1+1)+u_j+h]}{[h][(u_1+1)+u_j][u_j-(u_1+1)+1/2]}
=\frac{[1/2][u_j-u_1][u_1+u_j+h]}{[h][u_1+u_j][u_j-u_1+1/2]}, \\
&\frac{[1/2][u_j-(u_1-i \log ({\bf q})/\pi)][(u_1-i \log ({\bf q})/\pi)+u_j+h]}{[h][(u_1-i \log ({\bf q})/\pi)+u_j][u_j-(u_1-i \log ({\bf q})/\pi)+1/2]}
\nonumber \\
&=\exp (-2 \pi i (h+1/2) )\frac{[1/2][u_j-u_1][u_1+u_j+h]}{[h][u_1+u_j][u_j-u_1+1/2]},
\end{align}
and using the definition of the Pfaffian of a matrix
\eqref{definitionpfaffian},
one can calculate the quasi-periodicities of $e_{2}(u_1)=\mathrm{Pf} X$ and get
\begin{align}
&e_{2}(u_1+1)
=e_{2}(u_1), \label{qpcheckthree} \\
&e_{2}(u_1-i \log ({\bf q})/\pi)
=\exp (-2 \pi i (h+1/2) ) e_{2}(u_1). \label{qpcheckfour}
\end{align}
Combining \eqref{qpcheckone}, \eqref{qpchecktwo},
\eqref{qpcheckthree} and \eqref{qpcheckfour}, we get
the quasi-periodicities of $E_{2n}(u_1,\dots,u_{2n}|h)$
\begin{align}
&E_{2n}(u_1+1,\dots,u_{2n}|h)
=(-1)^{2n-1} E_{2n}(u_1,\dots,u_{2n}|h), \\
&E_{2n}(u_1-i \log ({\bf q})/\pi,\dots,u_{2n}|h)
 \nonumber\\
={}&(-{\bf q}^{-1})^{2n-1}
\exp \! \Bigg(-2 \pi i
\Bigg((2n-1)u_1+h+\sum_{j=2}^{2n} u_j
\Bigg)
\Bigg) E_{2n}(u_1,\dots,u_{2n}|h).
\end{align}
We can also check that
$E_{2n}(u_1+1,\dots,u_{2n}|h)$ is holomorphic as a function of $u_1$.
The factors $[u_i+u_j]$ and $[u_j-u_i+1/2]$
in the denominators of
$\displaystyle X_{ij}=
\frac{[1/2][u_j-u_i][u_i+u_j+h]}{[h][u_i+u_j][u_j-u_i+1/2]}$,
the matrix elements of $X$ which are used
to construct the Pfaffian,
are cancelled by the overall factor
$\displaystyle
\prod_{1 \le i < j \le 2n}
\frac{[u_j+u_i][u_j-u_i+1/2]}{[u_j-u_i]}$.
The factors $[u_k-u_1]$, $k=2,\dots,2n$ in the denominator
of the overall factor $\displaystyle
\prod_{1 \le i < j \le 2n}
\frac{[u_j+u_i][u_j-u_i+1/2]}{[u_j-u_i]}$
may lead to singularities at $u_1=u_k \ (k=2,\dots,2n)$, but do not.
For example, let us see the case $k=2$.
If one expands the Pfaffian of the matrix $X$, all summands containing $X_{12}$
as a factor of the product have the factor $[u_2-u_1]$ in the numerator.
The sum of the summands which do not contain $X_{12}$ can be rearranged
as a linear combination of the terms
$(X_{1j}X_{2k}-X_{1k}X_{2j})\prod_{\ell=3}^{n}X_{\sigma^\prime(2\ell-1) \sigma^\prime(2\ell)} \ (j,k \neq 1,2, \ \sigma^\prime \in M_{2n-4})$,
all of which vanish at $u_1=u_2$.
Hence, we find $E_{2n}(u_1,\dots,u_{2n}|h)$ is an elliptic polynomial
of degree $2n-1$ and we find Property (1) holds.

Next, let us show Property (2).
Applying the expansion formula for the Pfaffian \eqref{pfaffianexpansion}
to the matrix $X$, $E_{2n}(u_1,\dots,u_{2n}|h)$
can be expanded as
\begin{align}
&E_{2n}(u_1,\dots,u_{2n}|h) \nonumber \\
=&\prod_{1 \le i < j \le 2n}
\frac{[u_j+u_i][u_j-u_i+1/2]}{[u_j-u_i]}
\sum_{k=2}^{2n}
\frac{[1/2][u_k-u_1][u_1+u_k+h]}{[h][u_1+u_k][u_k-u_1+1/2]}
\mathrm{Pf} X_{1,k}^{1,k}. \label{ellipticpfaffianexpansion}
\end{align}
Then one notes that if one substitutes $u_1=-u_\ell$,
only the summand $k=\ell$ of the sum in
the right hand side of \eqref{ellipticpfaffianexpansion} survives.
Here, we use the basic property for the theta function $[0]=0$ for this observation.
After the substitution $u_1=-u_\ell$ in \eqref{ellipticpfaffianexpansion}
and after simplifications, one finds
\begin{align}
&E_{2n}(u_1,\dots,u_{2n}|h)|_{u_1=-u_\ell} \nonumber \\
=&[1/2]
\prod_{ \substack{ j=2 \\ j \neq \ell} }^{2n}[u_j+u_\ell+1/2][u_j-u_\ell+1/2]
\prod_{\substack{1 \le i < j \le 2n \\  i,j \neq 1,\ell}}
\frac{[u_j+u_i][u_j-u_i+1/2]}{[u_j-u_i]}
\mathrm{Pf} X_{1,\ell}^{1,\ell},
\nonumber \\
={}&[1/2]
\prod_{ \substack{ j=2 \\ j \neq \ell} }^{2n}[u_j+u_\ell+1/2][u_j-u_\ell+1/2]
E_{2n-2}(u_2,\dots,\hat{u_\ell} ,\dots,u_{2n}|h). \label{tomatch}
\end{align}
Here, we have used the fact that $[u]$ is odd $[-u]=-[u]$
and the property \eqref{usethisproperty} to get
the expression \eqref{tomatch} from
the expansion \eqref{ellipticpfaffianexpansion}.
Hence Property (2) is proved.

The only thing left to do is to check Property (3), which
can be easily seen from the definition
of $E_{2n}(u_1,\dots,u_{2n}|h)$ \eqref{definitionofE}.
\end{proof}

We can make another Proposition (Korepin's characterization
of the partition functions) which looks almost the same,
but is slightly different from Proposition
\ref{ordinarypropertiesfordomainwallboundarypartitionfunction},
i.e., a different version of the
elliptic Izergin-Korepin analysis which is presented below.

\begin{proposition} 
\label{anotherizerginkorepin}
The partition functions with OS boundary
$P_{2n}(u_1,\dots,u_{2n}|h)$
satisfy Properties (1) and (3) of
Proposition
\ref{ordinarypropertiesfordomainwallboundarypartitionfunction}, together with:
\\
The following relations among the
partition functions hold (Fig. \ref{picturerecursiontwo}):
\begin{align}
&\hspace{-36pt}P_{2n}(u_1,\dots,u_{2n}|h)
|_{u_1=-u_\ell-1/2}
\nonumber \\
={}&\frac{[h-1/2][1/2]}{[h]}
\prod_{ \substack{ j=2 \\ j \neq \ell} }^{2n}[u_j+u_\ell][u_j-u_\ell-1/2]
P_{2n-2}(u_2,\dots,\hat{u_\ell} ,\dots,u_{2n}|h), \label{anotherrecursion}
\end{align}
for $\ell=2,\dots,2n$,
and $\hat{u_\ell}$ in $P_{2n-2}(u_2,\dots,\hat{u_\ell} ,\dots,u_{2n}|h)$
means that $\hat{u_\ell}$ is removed.
\end{proposition}

\begin{proof}
The only additional thing to prove is \eqref{anotherrecursion}.
It is enough to prove the case $\ell=2n$
since the other cases $\ell=2,\dots,2n-1$
follow from the case $\ell=2n$ of \eqref{anotherrecursion}
by using the symmetry of $P_{2n}(u_1,\dots,u_{2n}|h)$
with respect to the variables $u_1,\dots,u_{2n}$.

We again use the graphical representation of $P_{2n}(u_1,\dots,u_{2n}|h)$
(Fig. \ref{picturerecursiontwo}).
We first realize that when we set $u_1$ to $u_1=-u_{2n}-1/2$,
the $R$-matrix at the southeast corner starts to freeze
since ${}_1 \langle 0 | {}_{2n} \langle 0 | R_{1,2n}(-u_{2n}-1/2,-u_{2n},h)|0 \rangle_1 | 0 \rangle_{2n}=0$,
and continuing graphical observation using the ice-rule of the $R$-matrix
\eqref{icerule}, one finds that the $R$- and $K$-matrices at the bottom row
and the rightmost column are frozen. The contribution
of these frozen parts to the partition functions is
the overall factor
$\displaystyle [h-1/2][1/2][h]^{-1}
\prod_{j=2}^{2n-1}[u_j+u_{2n}][u_j-u_{2n}-1/2]
$, and the remaining unfrozen part is
$P_{2n-2}(u_2,\dots,u_{2n-1}|h)$.
Hence, we get
\begin{align}
&P_{2n}(u_1,\dots,u_{2n}|h)|_{u_1=-u_{2n}-1/2} \nonumber \\
=&\frac{[h-1/2][1/2]}{[h]}
\prod_{j=2}^{2n-1}[u_j+u_{2n}][u_j-u_{2n}-1/2]
P_{2n-2}(u_2,\dots,u_{2n-1}|h).
\end{align}
\end{proof}

One can obtain an elliptic Pfaffian representation
of the partition functions which is similar to that of
\eqref{ellipticrepresentation}
in Theorem \ref{maintheorem} by finding a representation
satisfying the properties in Proposition \ref{anotherizerginkorepin}.
\begin{theorem} \label{maintheoremtwo}
The partition functions $P_{2n}(u_1,\dots,u_{2n}|h)$
with OS boundary have the following expressions
using elliptic Pfaffians:
\begin{align}
&P_{2n}(u_1,\dots,u_{2n}|h) \nonumber \\
=&\prod_{1 \le i < j \le 2n}
\frac{[u_j+u_i+1/2][u_j-u_i+1/2]}{[u_j-u_i]}
\mathrm{Pf} \Bigg( 
\frac{[1/2][u_j-u_i][u_i+u_j+h]}{[h][u_i+u_j+1/2][u_j-u_i+1/2]}
\Bigg)_{1 \le i,j \le 2n}. \label{anotherellipticrepresentation}
\end{align}
\end{theorem}

\begin{proof}
This can be proved in the same way as
proving Theorem \ref{maintheorem}.
For example, one can show that the RHS of \eqref{anotherellipticrepresentation}
satisfies \eqref{anotherrecursion} in the same way as
proving that the RHS of \eqref{ellipticrepresentation}
satisfies \eqref{ordinaryrecursionwavefunction}
using the expansion formula for the Pfaffian \eqref{pfaffianexpansion}.
\end{proof}

We derived two elliptic Pfaffian representations
of the partition functions $P_{2n}(u_1,\dots,u_{2n}|h)$
\eqref{ellipticrepresentation} in
Theorem \ref{maintheorem} and
\eqref{anotherellipticrepresentation} in
Theorem \ref{maintheoremtwo}
based on two versions of the Korepin's method
Proposition
\ref{ordinarypropertiesfordomainwallboundarypartitionfunction}
and Proposition \ref{anotherizerginkorepin}.
By comparing \eqref{ellipticrepresentation}
and \eqref{anotherellipticrepresentation},
we get the following identity between two elliptic Pfaffians.

\begin{theorem} \label{identitytheorem}
The following identity between two elliptic Pfaffians holds:
\begin{align}
&\prod_{1 \le i < j \le 2n}[u_j+u_i]
\mathrm{Pf} \Bigg( 
\frac{[u_j-u_i][u_i+u_j+h]}{[u_i+u_j][u_j-u_i+1/2]}
\Bigg)_{1 \le i,j \le 2n} \nonumber \\
=&\prod_{1 \le i < j \le 2n}[u_j+u_i+1/2]
\mathrm{Pf} \Bigg( 
\frac{[u_j-u_i][u_i+u_j+h]}{[u_i+u_j+1/2][u_j-u_i+1/2]}
\Bigg)_{1 \le i,j \le 2n}. \label{ellipticidentity}
\end{align}
\end{theorem}

The special case $h=0$ of the identity \eqref{ellipticidentity}
can be obtained by combining the following two factorization formulas
for the elliptic Pfaffians by
Rosengren \cite{Roselldet,Rosellpfaffian} and Rains \cite{Rains}.
\begin{align}
\mathrm{Pf} \Bigg( 
\frac{[u_j-u_i]}{[u_j-u_i+1/2]}
\Bigg)_{1 \le i,j \le 2n}&=
\prod_{1 \le i < j \le 2n}\frac{[u_j-u_i]}{[u_j-u_i+1/2]},
\label{Pfaffianfactorizationone} \\
\mathrm{Pf} \Bigg( 
\frac{[u_j-u_i][u_i+u_j]}{[u_i+u_j+1/2][u_j-u_i+1/2]}
\Bigg)_{1 \le i,j \le 2n}&=
\prod_{1 \le i < j \le 2n}\frac{[u_j-u_i][u_i+u_j]}
{[u_i+u_j+1/2][u_j-u_i+1/2]}
\label{Pfaffianfactorizationtwo}.
\end{align}

See Remark 2.1 in \cite{Roselldet} for example where it is explained that
\eqref{Pfaffianfactorizationone} is a special case
of the Pfaffian evaluation by Rosengren ((2.9) and (2.11) in \cite{Roselldet}),
and \eqref{Pfaffianfactorizationtwo} 
is a modular dual of the Pfaffian evaluation by Rains
(the last equation in Remark 2.1 in \cite{Roselldet} and
Theorem 2.10 in \cite{Rains}).

We directly check \eqref{ellipticidentity} for the case $n=2$
in the Appendix
by repeatedly using addition formulas for the theta functions \eqref{additionformula}.

\section{Conclusion}\label{sec5}
In this paper, we studied the partition functions
of the elliptic free-fermionic face model with OS boundary,
and analyzed them by using the elliptic Izergin-Korepin analysis.
We obtained the representations of the partition functions
using elliptic Pfaffians.
Since we can use the Korepin's method in two ways,
we can get two Pfaffian representations for the same partition functions.
As a corollary of the two expressions,
we get an identity between two elliptic Pfaffians.

It would be interesting to extend the analysis performed
on the OS boundary in this paper to other boundary conditions,
i.e., consider various variations
of the domain wall boundary partition functions
introduced by Kuperberg \cite{Ku2} for the case of the elliptic face models.
More complicated boundary conditions may lead to
expressions as products of determinants and Pfaffians
as is the case for the trigonometric models,
and may also lead to various interesting identities
between elliptic determinants and elliptic Pfaffians.
It would also be interesting to investigate if the
elliptic Pfaffian identities by
Okada \cite{Okadapfaffian},
Rosengren \cite{Roselldet,Rosellpfaffian} and Rains \cite{Rains}
can be understood as different representations of
the same partition functions of elliptic integrable models.

\section*{Acknowledgments}
The author thanks the referees for careful reading,
various invaluable comments and suggestions to improve the paper.
This work was partially supported by Grant-in-Aid
for Scientific Research (C)
No. 18K03205 and No. 16K05468.

\appendix

\section{Appendix: An elemenary proof of \eqref{ellipticidentity} for the case $n=2$}
In this Appendix, we check \eqref{ellipticidentity} for the case $n=2$
by elementary manipulations. In this case, one can see from the definition of
Pfaffians \eqref{definitionpfaffian}
that proving \eqref{ellipticidentity}
is equivalent to showing the following identity
\begin{align}
&\frac{[u_2-u_1][u_1+u_2+h][u_4-u_3][u_3+u_4+h]}{[u_2-u_1+1/2][u_4-u_3+1/2]} \nonumber \\
&\times[u_3+u_1+1/2][u_4+u_1+1/2][u_3+u_2+1/2][u_4+u_2+1/2] \nonumber \\
-&\frac{[u_3-u_1][u_1+u_3+h][u_4-u_2][u_2+u_4+h]}{[u_3-u_1+1/2][u_4-u_2+1/2]} \nonumber \\
&\times[u_2+u_1+1/2][u_4+u_1+1/2][u_3+u_2+1/2][u_4+u_3+1/2] \nonumber \\
+&\frac{[u_4-u_1][u_1+u_4+h][u_3-u_2][u_2+u_3+h]}{[u_4-u_1+1/2][u_3-u_2+1/2]} \nonumber \\
&\times[u_2+u_1+1/2][u_3+u_1+1/2][u_4+u_2+1/2][u_4+u_3+1/2] \nonumber \\
=&
\frac{[u_2-u_1][u_1+u_2+h][u_4-u_3][u_3+u_4+h][u_3+u_1][u_4+u_1][u_3+u_2][u_4+u_2]}{[u_2-u_1+1/2][u_4-u_3+1/2]}
\nonumber \\
-&\frac{[u_3-u_1][u_1+u_3+h][u_4-u_2][u_2+u_4+h][u_2+u_1][u_4+u_1][u_3+u_2][u_4+u_3]}{[u_3-u_1+1/2][u_4-u_2+1/2]} \nonumber \\
+&\frac{[u_4-u_1][u_1+u_4+h][u_3-u_2][u_2+u_3+h][u_2+u_1][u_3+u_1][u_4+u_2][u_4+u_3]}{[u_4-u_1+1/2][u_3-u_2+1/2]}. \label{identitytoshow}
\end{align}
Let us show this using the addition formula for the theta functions
\eqref{additionformula} repeatedly.
The difference between the left hand side and the right hand side of
\eqref{identitytoshow} can be expressed as
\begin{align}
&\frac{[u_2-u_1][u_1+u_2+h][u_4-u_3][u_3+u_4+h]}{[u_2-u_1+1/2][u_4-u_3+1/2]} \nonumber \\
\times&([u_3+u_1+1/2][u_4+u_1+1/2][u_3+u_2+1/2][u_4+u_2+1/2] \nonumber \\
&-[u_3+u_1][u_4+u_1][u_3+u_2][u_4+u_2])
\nonumber \\
-&\frac{[u_3-u_1][u_1+u_3+h][u_4-u_2][u_2+u_4+h]}{[u_3-u_1+1/2][u_4-u_2+1/2]} \nonumber \\
\times&([u_2+u_1+1/2][u_4+u_1+1/2][u_3+u_2+1/2][u_4+u_3+1/2]
\nonumber \\
&-[u_2+u_1][u_4+u_1][u_3+u_2][u_4+u_3])
\nonumber \\
+&\frac{[u_4-u_1][u_1+u_4+h][u_3-u_2][u_2+u_3+h]}{[u_4-u_1+1/2][u_3-u_2+1/2]}
\nonumber \\
\times&([u_2+u_1+1/2][u_3+u_1+1/2][u_4+u_2+1/2][u_4+u_3+1/2] \nonumber \\
&-[u_2+u_1][u_3+u_1][u_4+u_2][u_4+u_3]). \label{identitytoshowtwo}
\end{align}
Using the addition formula for the theta functions
\eqref{additionformula} (and \eqref{usethisproperty}),
one finds
\begin{align}
&[u_3+u_1+1/2][u_4+u_1+1/2][u_3+u_2+1/2][u_4+u_2+1/2] \nonumber \\
&-[u_3+u_1][u_4+u_1][u_3+u_2][u_4+u_2] \nonumber \\
=&[1/2][u_1+u_2+u_3+u_4+1/2][u_2-u_1+1/2][u_4-u_3+1/2], \label{addone} \\
&[u_2+u_1+1/2][u_4+u_1+1/2][u_3+u_2+1/2][u_4+u_3+1/2] \nonumber \\
&-[u_2+u_1][u_4+u_1][u_3+u_2][u_4+u_3]
\nonumber \\
=&[1/2][u_1+u_2+u_3+u_4+1/2][u_3-u_1+1/2][u_4-u_2+1/2], \label{addtwo} \\
&[u_2+u_1+1/2][u_3+u_1+1/2][u_4+u_2+1/2][u_4+u_3+1/2] \nonumber \\
&-[u_2+u_1][u_3+u_1][u_4+u_2][u_4+u_3] \nonumber \\
=&[1/2][u_1+u_2+u_3+u_4+1/2][u_3-u_2+1/2][u_4-u_1+1/2]. \label{addthree}
\end{align}
Using the identities \eqref{addone}, \eqref{addtwo} and \eqref{addthree},
\eqref{identitytoshowtwo} reduces to
\begin{align}
&\frac{[u_2-u_1][u_1+u_2+h][u_4-u_3][u_3+u_4+h]}{[u_2-u_1+1/2][u_4-u_3+1/2]}
\nonumber \\
\times&[1/2][u_1+u_2+u_3+u_4+1/2][u_2-u_1+1/2][u_4-u_3+1/2]
\nonumber \\
-&\frac{[u_3-u_1][u_1+u_3+h][u_4-u_2][u_2+u_4+h]}{[u_3-u_1+1/2][u_4-u_2+1/2]}
\nonumber \\
\times&[1/2][u_1+u_2+u_3+u_4+1/2][u_3-u_1+1/2][u_4-u_2+1/2]
\nonumber \\
+&\frac{[u_4-u_1][u_1+u_4+h][u_3-u_2][u_2+u_3+h]}{[u_4-u_1+1/2][u_3-u_2+1/2]}
\nonumber \\
\times&[1/2][u_1+u_2+u_3+u_4+1/2][u_3-u_2+1/2][u_4-u_1+1/2] \nonumber \\
=&[1/2][u_1+u_2+u_3+u_4+1/2] ([u_2-u_1][u_1+u_2+h][u_4-u_3][u_3+u_4+h]
\nonumber \\
-&[u_3-u_1][u_1+u_3+h][u_4-u_2][u_2+u_4+h]
+[u_4-u_1][u_1+u_4+h][u_3-u_2][u_2+u_3+h]). \label{finalstep}
\end{align}
One can apply
the addition formula \eqref{additionformula} again to get
\begin{align}
&[u_2-u_1][u_1+u_2+h][u_4-u_3][u_3+u_4+h]
-[u_3-u_1][u_1+u_3+h][u_4-u_2][u_2+u_4+h] \nonumber \\
+&[u_4-u_1][u_1+u_4+h][u_3-u_2][u_2+u_3+h]=0,
\end{align}
and we find the right hand side of \eqref{finalstep} becomes zero.
Hence \eqref{ellipticidentity} for the case $n=2$
is proved.

\end{document}